# A Lunar L2-Farside Exploration and Science Mission Concept with the Orion Multi-Purpose Crew Vehicle and a Teleoperated Lander/Rover

Jack O. Burns[a,b,*], David. A. Kring[c,b], Joshua B. Hopkins[d], Scott Norris[d], T. Joseph W. Lazio[e,b], Justin Kasper[f,b]

[a]Center for Astrophysics and Space Astronomy, Department of Astrophysical and Planetary Sciences, 593 UCB, University of Colorado Boulder, Boulder, CO 80309, USA
[b]NASA Lunar Science Institute, NASA Ames Research Center, Moffett Field, CA 94089, USA
[c]Center for Lunar Science and Exploration, USRA Lunar and Planetary Institute, 3600 Bay Area Blvd., Houston, TX 77058 USA
[d]Lockheed Martin Space Systems, P.O. Box 179, CO/TSB, M/S B3004, Denver CO 80127 USA
[e]Jet Propulsion Laboratory, California Institute of Technology, MS 138-308, 4800 Oak Grove Dr., Pasadena, CA 91109, USA
[f]Harvard-Smithsonian Center for Astrophysics, Perkins 138, MS 58, 60 Garden St., Cambridge, MA 02138, USA

**Abstract**

A novel concept is presented in this paper for a human mission to the lunar L2 (Lagrange) point that would be a proving ground for future exploration missions to deep space while also overseeing scientifically important investigations. In an L2 halo orbit above the lunar farside, the astronauts aboard the Orion Crew Vehicle would travel 15% farther from Earth than did the Apollo astronauts and spend almost three times longer in deep space. Such a mission would serve as a first step beyond low Earth orbit and prove out operational spaceflight capabilities such as life support, communication, high speed re-entry, and radiation protection prior to more difficult human exploration missions. On this proposed mission, the crew would teleoperate landers and rovers on the unexplored lunar farside, which would obtain samples from the geologically interesting farside and deploy a low radio frequency telescope. Sampling the South Pole-Aitkin basin, one of the oldest impact basins in the solar system, is a key science objective of the 2011 Planetary Science Decadal Survey. Observations at low radio frequencies to track the effects of the Universe's first stars/galaxies on the intergalactic medium are a priority of the 2010 Astronomy and Astrophysics Decadal Survey. Such telerobotic oversight would also demonstrate capability for human and robotic cooperation on future, more complex deep space missions such as exploring Mars.

[*] Corresponding author at: Center for Astrophysics and Space Astronomy, Department of Astrophysical and Planetary Sciences, 593 UCB, University of Colorado Boulder, Boulder, CO 80309, USA.
   *E-mail address:* jack.burns@colorado.edu (J. O. Burns).

# 1. Introduction

The Moon's farside is a possible early goal for missions beyond Low Earth Orbit (LEO) using NASA's Orion Multi-Purpose Crew Vehicle (MPCV) in tandem with teleoperated robots to explore incrementally more distant destinations. The lunar L2 Lagrange Point is a location where the combined gravity of the Earth and Moon allows a spacecraft to be synchronized with the Moon in its orbit around the Earth, so that the spacecraft is relatively stationary over the farside of the Moon (Fig. 1).

The farside has been mapped from orbit, but no humans or robots have ever landed there. There are two important science objectives on the farside. The first would be to return to Earth multiple rock samples from the Moon's South Pole–Aitken (SPA) basin, one of the largest, deepest, and oldest impact basins in the solar system. A sample return from SPA was designated as a priority science objective in the NRC Decadal Survey *Vision and Voyages for Planetary Science in the Decade 2013-2022* (NRC, 2011) as well as the NRC report *The Scientific Context for Exploration of the Moon* (NRC, 2007). The second objective would be to deploy a low radio frequency telescope, where it would be shielded from human-generated radio frequency interference (RFI) from the Earth and free from ionospheric attenuation, allowing astronomers to explore the currently unobserved *Dark Ages* and *Cosmic Dawn* of the early Universe. These observations were identified as one of the top science objectives in the NRC Decadal Survey *New Worlds, New Horizons in Astronomy and Astrophysics* (NRC, 2010): "Cosmic Dawn: Searching for the First Stars, Galaxies, and Black Holes," as well as a science priority in the NRC (2007) report. Thus, this proposed mission would fulfill some of the top science goals in planetary science *and* in astrophysics as proposed in recent NRC reports.

A robotic lander and rover would be launched first on a slow but efficient Weak Stability Boundary trajectory to the Moon (also known as a ballistic lunar transfer), to ensure that the rover is on its way before the crew is launched. Next, three astronauts would be launched in an Orion spacecraft using NASA's heavy-lift Space Launch System (SLS). Orion would fly past the Moon for a gravity slingshot maneuver towards the L2 point. Orion would use its propulsion system to enter a halo or Lissajous orbit around the L2 point. From this vantage point, ≈65,000 km above the farside of the Moon, Orion would have continuous line-of-sight visibility to both the farside of the Moon and the Earth (Fig. 2).

The proposed science objectives do not require teleoperation of a rover from L2 since this science might be better accomplished with the assistance of a human crew on the surface or less effectively via control from Earth using a relay satellite at L2. However, by using teleoperation of rovers by astronauts at L2, this mission would demonstrate human "virtual presence" from orbit to explore and deploy sophisticated science instrumentation on an extraterrestrial body (see also Lester and Thronson 2011b). The two-way speed of light latency is only 0.4 seconds between L2 and the farside surface so this will permit real-time control of the rover. This mission will provide crucial operational experience for more challenging destinations such as permanently-shadowed <40 K craters at the lunar poles, recently shown to possess water ice, and potential biologically-interesting sites on Mars.

This paper begins with a more detailed overview of the Orion MPCV L2-Farside mission concept in Section 2. In Section 3, a landing site is proposed for the robotic lander/rover on the farside – the Schrödinger crater that lies within the South Pole-Aitken basin. The scientific advantages of this young impact basin are presented and the geological telerobotic exploration goals of the Schrödinger crater are discussed. In Section 4, the deployment of a roll-out polyimide, low frequency antenna array is described along with the science goals that include tracking the effects of the first stars and galaxies via redshifted 21-cm radiation from the neutral hydrogen in the intergalactic medium. Secondary science with this array include potential measurements of the lunar ionosphere and of interplanetary nanoscale dust. A summary of this mission concept is given in Section 5.

## 2. The Orion MPCV L2-Farside Mission Concept

*2.1 Overview*

The Orion Multi-Purpose Crew Vehicle (MPCV) spacecraft has core capabilities that will enable the L2-Farside mission since Orion was designed from its inception to support lunar missions.

Orion consists of four major elements, shown in Fig. 3. The Launch Abort System (LAS) at the top is designed to pull the spacecraft away from launch vehicle in the event of an emergency during ascent. The conical Crew Module (CM) contains the pressurized living space for the astronauts, as well as most of the vehicle avionics. The CM returns the crew to Earth, using a heat shield and parachutes. The CM also has a docking adapter at the top to connect to other spacecraft. Below the CM is the Service Module (SM), which provides most of the utility functions on the spacecraft. It contains the propellant tanks and main engine for propulsion, the tanks of water and oxygen for life support, solar arrays for power, a thermal control fluid loop with radiators to cool the capsule, and a phased array antenna for long distance communication. A Spacecraft Adapter connects Orion to its launch vehicle. External jettisoned panels cover the solar arrays, radiators, and thrusters during ascent.

Orion has all the necessary capabilities to operate in deep space and for a long duration, such as solar arrays for power generation, regenerative amine beds rather than single-use lithium hydroxide canisters to remove $CO_2$ from the cabin atmosphere, and the design robustness necessary to ensure the vehicle withstands Micro-Meteoroids and Orbital Debris (MMOD) impacts. These capabilities allow Orion to meet the needs of lunar missions and provide sufficient performance for the lunar farside mission. Table 1 highlights several of the Environmental Control and Life Support Systems (ECLSS) along with the technology that supports the function.

*2.2 Mission Concept*

The L2-Farside mission concept would send astronauts in the Orion to a halo or Lissajous orbit around the second Earth-Moon Lagrange point, a location where the spacecraft appears to hover over the farside of the Moon (Fig. 1). Prior to this proposed mission, the Orion will undergo several development tests to ensure crew safety and mission viability. The first Exploration Flight Test (EFT-1) is scheduled for 2014 to verify the crew module can survive high speed re-entry into the Earth's atmosphere from a Lunar return trajectory. Three additional test flights are planned for the Orion and include a high altitude abort, an un-crewed flight around the Moon in 2017, and a crewed flight soon after that. Because Orion is designed for lunar missions, only a few minor modifications are required to operate at L2, primarily to increase the life support consumables supply to enable at least a one month mission duration capability and to add control equipment to teleoperate lunar rovers. It may be possible to integrate rover control capabilities into Orion's existing programmable displays and controls, but a more likely approach is to bring dedicated laptop-type control stations. Functional requirements for other aspects of the L2 mission such as reliable in-space propulsion, re-entry capability for the appropriate velocity, multi-channel high-gain communications, and thermal management are satisfied by the existing design.

The launch vehicle for this proposed mission to explore the Moon's farside would be the Block 1 configuration of the Space Launch System (SLS), a heavy lift launch vehicle currently under development by NASA. A robotic lander and rover would be launched first using a separate, smaller launch vehicle such as an Atlas V. Next, three astronauts would be launched in the Orion MPCV on SLS (Fig. 4). Orion is designed to carry up to four astronauts, but reducing the crew to three for this mission provides each person with more living space in the CM. Orion would use a technique identified by Farquhar (1970), which uses a lunar flyby trajectory to reach L2. The Orion MPCV would fly past the Moon at a perilune altitude on the order of 100 km for a propulsive gravity slingshot maneuver towards the L2 point, where it would use its propulsion system to enter into a halo orbit around the Lagrange point. The total transfer time to L2 using a lunar swingby is a few days longer than a direct trajectory, but the $\Delta V$ savings of several hundred m/s are substantial enough to justify the additional trip time. We have developed reference trajectories with daily launch windows that have a total transfer time from Earth to L2 of 8-12 days with a round-trip spacecraft $\Delta V$ budget of under 500 m/s after Translunar Injection (TLI).

Around the L2 point, approximately 65,000 km above the farside of the Moon, the Orion would have continuous line-of-sight communications access to both the lunar farside and the Earth. Astronauts could operate a rover on the lunar surface and stay in contact with mission control at the same time, orbiting the L2 point for about two weeks—long enough to operate a rover through the full length of a lunar daylight cycle (see also Bobskill and Lupisella 2012).

The Moon is close enough to Earth that rovers could be controlled from Earth using a simple relay satellite, but the short speed of light lag (≈0.4 s) makes real-time control from L2 viable. According to Lester and Thronson (2011a), the "cognitive horizon" for real-time telepresence is ≈0.5 s. Hopkins (2012) reported on tests using a simulated lunar rover control interface with artificially introduced latency which indicated that operators faced with latencies similar to controlling from L2 could drive roughly twice as fast as when the latency was equivalent to Earth-based control. However, the actual difference in efficiency will depend on the complexity of the operations in question

Controlling the rovers from an orbiting spacecraft like Orion would be useful practice for future Mars missions. The 2009 Augustine Committee endorsed an idea long advocated by others (e.g. Singer, 1984 and Landis, 1995), that the first human missions to Mars should not attempt the difficult step of landing on the surface, but may instead orbit Mars and control robot rovers on the surface (Kwong et al. 2011, GSFC 2012). This approach is likely to be more effective than current methods of controlling the rovers from Earth, because Mars is so far away that the speed-of-light delay and gaps in communications coverage slow down operations so much that simple operations can take many days.  Schmidt, Landis and Oleson have elaborated on this idea in their HERRO study and point out that joint human/robotic exploration via telepresence can also be used to explore destinations which are inhospitable to humans such as Venus (Schmidt et al, 2011). Mission operations using a telepresence approach to exploration will be very different from how either traditional astronaut spaceflights or robotic science missions have been operated in the past. The lunar L2-Farside missions would develop, implement and evaluate different operational methods for this type of joint human/robot exploration. In particular, it would address how planning, control, and decision-making functions should be divided between rover autonomy software, the few 'on-site' astronauts in a spacecraft, and the larger team of engineers and scientists in a mission control center on Earth. These early missions would also provide feedback on how much bandwidth is required, what types of sensors are most useful, and on how best to design astronaut control interfaces for surface telerobotics.

*2.3 Mission Duration*

The Orion was designed to NASA long duration human space flight architecture requirements, which specified that the vehicle had to support four astronauts for 21 days going to and from the Moon, with a 180-day unoccupied period in lunar orbit, plus 30 days of contingency loiter capability for a mission extension. Although the standard crewed duration of 21 days is shorter than the 30-35 days required for this mission, it can be easily increased by adding one extra water tank and using larger diameter oxygen storage tanks. This provides enough supplies to get to L2, stay for a 14 day lunar daylight cycle, and return. If multi-month stays at L2 are necessary to complete objectives on the farside, more living space and supplies are required. Lockheed Martin has collaborated with international industry partners to investigate how derivatives of spacecraft currently used to supply the International Space Station could be delivered to L2. Orion could dock to one of these vehicles to take advantage of additional supplies and living space for longer missions (Hopkins et al. 2012).

Mission durations can also be constrained by radiation exposure safety limits for the crew. NASA's current estimates indicate that missions beyond Earth's protective magnetic field should be limited to no more than about six months to keep exposure to Galactic cosmic rays below an acceptable risk threshold (e.g., NRC 2008). The proposed mission duration of about one month is acceptable.

**3. Science Goal: The Schrödinger Impact Basin as a Preserve of Solar System Evolution**

Impact craters larger than 300 km on the Moon are called basins to highlight their vast size and dramatic topography.  Basins have broad flat floors, an uplifted peak ring, and a series of collapsed blocks of the Moon's crust that slumped inward from the basin rims during the final phase of their formation.  Schrödinger basin is the best preserved impact basin of its size on the Moon and provides tremendous science and exploration opportunities, safe landing zones, and a landscape that could be navigated by robotic assets.  The basin lies on the lunar farside, is centered at 75° south, 132.4° east, and is contained within the immense South Pole-Aitken basin (Fig. 5), which is a primary target for investigations outlined by the National Research Council (2007, 2011).  The 320 km diameter Schrodinger basin is ~4 km deep and hosts a 150 km diameter inner peak ring that rises up to 2.5 km above the basin floor. The east, west, and south sides of the basin are ~200, ~245, and nearly 310 km from the limb of the Moon.  A landing site on the relatively flat floor of the basin interior would be comfortably 350 to 450 km from the limb(s).

Schrödinger has been classified as Imbrian in age (≈3.8 billion years old) based on the number of craters superposed on the basin and its corresponding ejecta blanket (Shoemaker et al. 1994; Wilhelms et al. 1987). It is the second youngest basin and, thus, one of the last basins to have formed during the intense basin-forming impact period that re-shaped planetary surfaces throughout the inner solar system before 3.8 billion years ago (Fig. 6). Not only does Schrödinger reflect evolutionary processes occurring at that time, its basin walls and uplifted peak ring contain rock from older episodes in lunar history. In addition, long after the basin was produced, magmas rose beneath the crater floor and produced small eruptions of volcanic material that may be less than 1 billion years old. Thus, collectively, missions to Schrödinger basin would be able to access rock produced over a very broad fraction of the Moon's history.

*3.1 Lunar Surface Science Priorities*

We propose to place a lander/rover within the Schrödinger basin, where the first and second highest priorities in the NRC report *The Scientific Context for Exploration of the Moon* (NRC, 2007) and over half of the remaining goals in that report could be addressed. Because it lies within the South Pole-Aitken basin, it is also a featured target of the NRC Decadal report *Vision and Voyages for Planetary Science in the Decade 2013-2022* (NRC, 2011). The highest science priority of the NRC (2007) report is to test the lunar cataclysm hypothesis, which suggests the Earth and Moon were severely modified by a swarm of asteroids circa 3.9-4.0 billion years ago. The concept of an impact cataclysm emerged from analyses of samples collected by Apollo astronauts. All of those samples, however, were collected in a relatively small region of the Moon (representing less than 5% of the surface) and none of them provide any information about events on the lunar farside, which remains a completely unexplored region.

To evaluate that hypothesis, it is critical that the ages of basins be measured to determine the magnitude and duration of the basin-forming epoch. Among the most important targets is Schrödinger basin, which represents the type of impact activity that occurred immediately prior to the earliest evidence of life on Earth (Fig. 6).

The second highest priority in the NRC (2007) report is to determine the age of the oldest basin-forming event to anchor the beginning of the basin-forming epoch. The oldest (and largest) basin is the South Pole-Aitken basin (see also Duke, 2003; Joliff et al. 2010; Alkalai et al. 2013). Because Schrödinger (Fig. 7) formed within the South Pole-Aitken basin, samples of the latter will exist within Schrödinger. Thus, in one geologic terrain, samples for both the oldest and the second youngest basins could be collected, essentially bracketing the entire basin-forming epoch.

In addition to solving the two highest priority chronological problems, those same impact melt samples could be used to determine the source of the impact projectiles and their chemical compositions. Some models suggest comets are responsible, while others suggest asteroids are responsible. These data, which identifies the source of the impactors in the solar system, would, in turn, provide the information needed to test proposed mechanisms for the impact flux, some of which include dramatic shifts in the orbits of the giant planets. These data could also be used to calculate the delivery of biogenic elements and any environmental consequences produced by the bombardment, which have been tentatively linked to the origin and early evolution of life on Earth. If the sample collection system were mobile with sufficient range and speed (as might be accommodated by real-time telepresence), then sampling of melts from other craters and basin-forming events may also be possible.

Schrödinger basin would also be an excellent probe of the lunar interior. Normal faults in the modification zone of the basin expose subsurface lithologies and their stratigraphic relationships. The uplifted central peak ring exposes even deeper levels in the Moon's crust and directly taps rocks produced within the lunar magma ocean (Kramer et al. 2012; Yamamoto et al. 2012). Furthermore, clasts of subsurface lithologies were excavated and entrained in impact melt breccias deposited within the crater. Thus, by combining observations of the modification zone, central uplift, and impact breccias, one can generate a cross-section of the lunar crust several 10's of kilometers deep. The volume of material beneath the impact site that was melted extends nearly to the base of the Moon's crust. Because that melt was well-mixed, samples of it would provide an average chemical composition of the entire crust of the south polar region of the Moon. Consequently, while collecting samples to determine the impact flux to the lunar surface, one would also be collecting samples of the lunar interior.

The combination of impact and volcanic samples of different ages, particularly in concert with the Apollo samples, would provide the data needed to calibrate the crater counting chronology used to determine the ages of surfaces around the entire Moon.

*3.2 Landing Sites*

Specific landing sites within Schrödinger for human crew operations were previously studied to facilitate the objectives defined above (O'Sullivan et al. 2011). The best geology would be produced by a well-trained crew on the surface. However, in a restricted budgetary environment that cannot generate a lander for crew, telerobotic exploration by crew from an Orion platform provides an alternative way to address the objectives, while simultaneously developing the operational capability needed for other exploration objectives. To illustrate the feasibility of telerobotic surface operations, we step through those previously identified crew landing sites (Fig. 8). The Schrödinger basin is such an exciting place, however, that it is the focus of much new research (Mest 2011; Yamamoto et al. 2012; Kramer et al. 2012). Prior to launch, we will want to take advantage of these higher fidelity views of the interior of the basin to generate a new landing site assessment that features the capabilities of robotic assets.

The first proposed landing site explored occurs on the northern portion of the central melt sheet, which provides a relatively smooth surface for landing. A sample of the Schrödinger impact melt could be collected at the landing site and secured for return to Earth. From that location, a rover could access two mare-type volcanic deposits to determine the source of the lavas in the lunar interior and to provide an age of the eruptions. Faults and fractures could also be accessed. Those tectonic features and much younger small craters that puncture the melt sheet could be used to evaluate the amount of chemical and mineralogical differentiation that occurred. With mobility capabilities extending to 10 to 20 km, the rover could also access samples of the lunar magma ocean in the central peak ring. Small craters along the way were produced by blocks of rock ejected from the Orientale basin; if remnants of those blocks could be found, then samples of the far western portion of the Moon could also be collected.

The proposed second landing site is located on the central melt sheet in the western part of the inner basin. A rover deployed from this location could collect samples of that melt before exploring an intriguing ridge whose origins are unclear, all within 10 km. If the mobility capability is 20 km, then the rover could access the peak ring and its samples of the lunar magma ocean. Some of the small craters in this region appear to have been produced by debris from the Antoniadi crater; if any remnant debris is found, it would provide samples of yet another part of the Moon.

The third landing site is located on an impact melt-bearing breccia, rather than the central melt sheet. Thus, landing and surface operations may be more challenging here than at the other two sites. However, this landing site would provide access to a dramatic pyroclastic vent that distributed the types of materials targeted for future in-situ resource utilization purposes. That vent is located ~8 km from the proposed landing site. That vent is also very deep and produces a permanently shadowed region, thus providing an opportunity to study lunar volatile issues. Dramatic graben occur in the vicinity of the vent that could be used to study both the magmatic and tectonic history of the region.

### 3.3 Telerobotic Lunar Surface Operations

This mission concept is designed to produce high-quality science while testing operational ideas that would feed forward into future exploration activities elsewhere on the Moon, asteroids, the moons of Mars, and the surface of Mars. When the crew arrive on station in a halo orbit about the Earth-Moon L2 position, a lander with a rover would already have been deployed within Schrödinger basin. A pre-planned traverse with geologic stations would have also been developed by the mission's science team. From the L2 location, crew would then implement that traverse sequence using telerobotic commands from the Orion platform. Based on their observations during the traverse, crew would have the responsibility to modify the pre-mission plan to successfully accomplish mission objectives, in collaboration with the science team. All activities would be monitored in nearly real time by the Mission Operations Center and Science Operations Center in Houston, because Orion could relay the video and command sequences from the lunar farside in an E-M L2 spacecraft configuration.

The lunar science objectives of the surface activities would be to collect samples for return to Earth and document the geologic context of those samples. Because high precision analyses would be conducted on Earth, the amount of instrumentation on the rover could be minimized. An imaging system would be needed for navigation, station context, and sample documentation. That system should have a dynamic range for operations under a variety of lighting conditions, potentially including supplemental illumination for shadowed operations. Mechanical devices like an arm with grappling capability would be required for sample collection and potentially trenching. As the rover will also deploy radio frequency antennas, its capabilities could also include the deployment of a radiation monitor, particularly if the monitor could be buried below the regolith to assess its shielding capability for future long-duration human surface missions. Long term monitoring of surface conditions would be possible, as a communication relay asset would be left in the L2 location when the crew returns to Earth. Schrödinger basin is also sufficiently large and geologically interesting to provide multiple mission opportunities that would test a variety of operational scenarios while generating high-priority science results.

At the end of the traverse activities, the rover would return to the lander, transfer samples to the ascent vehicle, and then retreat to a safe stand-by distance from where its imaging systems could capture the ascent and monitor any dust pluming created by that ascent.

Earth-based telerobotics have previously been demonstrated with the Lunakhod on the Moon and Pathfinder, Spirit, and Opportunity on Mars. This mission would, however, be the first demonstration of telerobotic operations on a planetary surface from a space-based platform. The crew would need to have significant geologic training to conduct the operations (e.g., Kring 2010).

In summary, in-situ geologic reconnaissance with a return of samples to Earth from the Schrödinger basin could tackle the first and second highest priorities of the NRC (2007) report. Of the eight concepts identified in that report, seven of them could be addressed within the basin and over twenty of the reports objectives could be considered. For these reasons, the Schrödinger basin is arguably the highest priority landing site on the lunar surface.

**4. Science Goal: Tracking Cosmic Dawn with a Low Radio Frequency Telescope on the Lunar Farside**

*4.1 Polyimide Film Antennas Deployed on the Lunar Surface*

We have developed an innovative approach for the deployment of large numbers of radio antennas on the lunar surface using polyimide film as a backbone (Lazio et al. 2011 and references therein). Polyimide film is a flexible substance with a substantial heritage in space flight applications. In this concept, a conducting substance is deposited on the polyimide film to form the antenna. The film would be rolled for storage in a small volume during transport. Once on the Moon, the polyimide film would be unrolled to deploy the antennas. The antennas would then be electronically phased to produce a radio interferometer where the angular resolution ($\theta$) depends upon the wavelength ($\lambda$) and maximum distance between dipoles (D) such that $\theta \sim \lambda/D$.

The film would also contain the transmission line system for conducting electrical signals back to the central node at the intersection of the arms. A central processor would provide digitization and filtering, and then would downlink the data to the MPCV for transmission to Earth.

We have undertaken a series of tests of metal-coated polyimide films in the field and in the lab, for which Lazio et al. (2011) summarize the first results. First, a test was conducted to assess the electrical performance of such an antenna. The feed point impedance of a single polyimide film antenna lying directly on the ground has been measured as a function of a wide radio frequency range. Good agreement has been found with computer simulations, and, in one set of tests, absorption of the received power levels was consistent with that expected from the Earth's ionosphere. Second, using a vacuum chamber at U. Colorado, we simulated the lunar environment in which the polyimide film is to be placed. The chamber contained a UV lamp to simulate solar radiation on the Moon and the table upon which polyimide film was placed was thermally cycled between equivalent lunar day (100 C) and night (-150 C) temperatures. No significant stiction was detected in the cycled polyimide film. No changes in the tensile strength, electrical conductivity, or flexibility were measured beyond the 5% level after repeated thermal cycling and UV exposure indicating that the polyimide film is an excellent choice for low frequency antenna backbone material.

We propose to deploy the polyimide film antenna array using the same telerobotic rover that was described in Section 3.3. Fig. 9 illustrates our proposed rover deployment strategy. Most importantly, this would demonstrate the feasibility of telerobotic deployment of large structures at remote locations in the solar system including, in the future, Mars and asteroids. We are preparing to perform a terrestrial test in 2013 of this telerobotic deployment strategy of polyimide film using the K10 rover at the NASA Ames Research Center operated by crew aboard the International Space Station (Bualat et al. 2012, Fong et al. 2012).

*4.2 The Science Program for a Farside Low Frequency Radio Array*

*4.2.1 Cosmic Dawn – The First Stars and Galaxies in the Early Universe*
The *New Worlds, New Horizons in Astronomy and Astrophysics (NWNH) Decadal Survey* (NRC, 2010) identified "Cosmic Dawn" as one of the three science objectives guiding the science program for this decade. The Survey asked "What were the first objects to light up the Universe and when did they do it?" In other words, how and when did the first stars, galaxies, and quasars form in the early Universe leading to the rich structure that we observe today with observatories such as the Hubble Space Telescope? In the science program articulated in *NWNW*, the *Epoch of Reionization* (EoR) and *Cosmic Dawn* were identified as science frontier discovery areas that could provide the opportunity for "transformational comprehension, i.e., discovery."

Using the Moon as a platform for probing Cosmic Dawn via low radio frequency astronomy observations has been recognized in other reports and community documents. As recent examples, both of the NRC report *The Scientific Context for the Exploration of the Moon* (NRC, 2007) and *The Lunar Exploration Roadmap: Exploring the Moon in the 21$^{st}$ Century: Themes, Goals, Objectives, Investigations, and Priorities"* produced by the Lunar Exploration Analysis Group (LEAG, 2011) discuss the scientific value of a lunar radio telescope.

Burns et al. (2012) have developed a concept for a new cosmology mission that would be the first to explore the Cosmic Dawn epoch of the Universe. The science described by Burns et al. (2012) could be accomplished via low radio frequency antennas in lunar orbit or by the farside polyimide array discussed in 4.1.

The specific science objectives for the Cosmic Dawn observations include (1) When did the first stars form? (2) When did the first accreting black holes form? (3) When did Reionization begin? (4) What surprises does the end of the Dark Ages and the beginning of Cosmic Dawn hold (e.g., dark matter decay)? We propose to use the highly-redshifted hyperfine 21-cm transition[†] from neutral hydrogen to track the formation of the first luminous objects by their impact on the intergalactic medium (IGM) at redshifts 11 – 35 (80-420 million years after the Big Bang).

---

[†] The 21-cm spectral line arises from a "spin-flip" transition when angular momentum vectors for the proton and electron flip from parallel to anti-parallel.

The evolution of the intergalactic medium following the Big Bang and the formation of the first stars is predicted to have an effect on the temperature of neutral hydrogen, with the result that the spin-flip transition of neutral hydrogen with a rest wavelength of 21-cm (1420 MHz) may be detectable over the redshift range of about 7 to 100. This redshift range corresponds to frequencies between ~10 and 200 MHz. There are numerous ground-based projects focused on detecting the redshifted hydrogen signals at redshifts <10. These projects range from single antenna systems to designs for a large, international array of telescopes known as the Square Kilometre Array (SKA). The primary focus of these projects tends to be on redshifts of about 10 (frequency ~ 150 MHz), as a number of complicating effects are likely to be less severe at higher frequencies than at lower frequencies (e.g., Earth's ionosphere, magnitude of corrupting foreground signals). Further, there is wide-spread community acknowledgement that observations will require space-based observations to obtain the highest precision measurements and to cover the entire redshift range.

The measurement approach proposed here is to track the influence of the first stars, galaxies, and black holes on the neutral IGM by means of the spectrum produced by the hydrogen hyperfine transition.. Fig. 10 shows a model of the sky-averaged, redshifted 21-cm spectrum from the Dark Ages (before the first stars) into Cosmic Dawn and Reionization. The brightness of this signal evolves with redshift, with several key inflection or "Turning Points" indicated in Fig. 10. By measuring the frequencies of these Turning Points, we would determine (a) the redshifts of when the first stars ignited and galaxies formed, (b) when the first black holes turned on and began heating the IGM via X-rays from their accretion disks, and (c) when the Universe began its final phase transition in evolving from an all neutral hydrogen IGM during the Dark Ages to an all ionized IGM at the end of the EoR (Burns et al. 2012 and references therein).

At radio frequencies ≈1-100 MHz, the lunar farside is the only location in the inner solar system that is free of human-generated radio frequency interference (RFI) and also uninhibited by ionospheric effects. Thus, it is the ideal location for our proposed Dark Ages/Cosmic Dawn experiment. Both the Radio Astronomy Explorer-2 (RAE-2) (Alexander et al. 1975) and the Apollo Command Modules, which had RF systems at low radio frequencies, observed complete cessation of emissions from Earth when they passed into the radio-quiet zone above the lunar farside. This contrasts with observations from the ground or Earth orbit where both RFI and the Earth's ionosphere seriously interfere with astronomical observations (Fig. 11). Thus, the lunar farside is ideal for observations and experiments at radio low frequencies.

As a result of being physically shielded from the Earth, both the lunar farside and a volume above it are recognized by the International Telecommunications Union (ITU) as deserving of special protection for radio astronomy. ITU-R RA.479 describes a shielded zone of the Moon (SZM) and makes recommendations about the use of the radio spectrum within it. Frequencies below 30 MHz are recommended for the exclusive use of radio astronomy and frequencies between 30 and 300 MHz should not be used by "active services" (i.e., transmitters) with certain, limited exceptions. Notably, these frequencies recommended for protection for radio astronomical use are lower than typically used for communications, either between spacecraft or to and from space and ground.

Of more serious concern are "out of band" transmissions or unintended emissions---emissions resulting from equipment on spacecraft which could radiate power into the frequencies below 300

MHz. While there are standard procedures for ensuring that spacecraft emissions are maintained below some level, the potentially harmful interference levels for the kinds of radio astronomical measurements intended for the lunar farside are (far) more demanding than typical limits mandated for spacecraft and will require attention to spacecraft design to ensure that the desired data acquisition can occur from lunar farside radio antennas.

We have investigated the electromagnetic (EM) environment of the Schrödinger basin on the lunar farside by running an EM wave propagation simulation tool based on a finite-difference time-domain (FDTD) method (Takahashi 2002). We examined low frequency diffraction effects from Earth-based RFI around the Moon's limb and in the crater as well as the impact of the electrical properties of the lunar surface. The initial results from our simulations indicate that incoming RFI EM waves experience >80 dB attenuation before they reach the crater interior. In order to detect faint cosmological signals described above, this attenuation is sufficient to establish the Schrödinger basin as a potential location for the proposed radio array. Currently, we are further developing this existing toolbox to carry out detailed, high resolution simulations of the lunar farside for incident EM waves with $\nu \geq 10$ MHz.

### 4.2.2 *Measurement of the Lunar Ionized Atmosphere*

The lunar atmosphere is the exemplar and nearest case of a surface boundary exosphere for an airless body in the solar system, and the recent *Planetary Science Decadal Survey* (NRC, 2011) noted the importance of tracking the evolution of exospheres, particularly in response to the space environment. Determining and tracking the properties of the lunar atmosphere both robustly and over time requires a lunar-based methodology by which the atmosphere could be monitored over multiple day-night cycles from a fixed location(s), such as a lunar relative ionosphere opacity meter (*riometer*).

Reviews of the state of knowledge of the lunar atmosphere at the close of the Apollo era indicate significant advances in knowledge of the composition, sources and sinks, and influences on the lunar atmosphere but also significant questions about all of these topics (e.g. Johnson 1971; Hodges et al. 1974). Exposed to both the solar and interstellar radiation fields, the daytime lunar atmosphere is mostly ionized. Enduring questions include the density and vertical extent of the ionosphere and its behaviour over time, including modification by robotic or crewed landers.

ALSEP measurements during the Apollo missions found a photoelectron layer near the surface with electron densities up to $10^4$ cm$^{-3}$ (Reasoner and O'Brien 1972). Further, dual-frequency radio occultation measurements from the Soviet Luna spacecraft suggest that the ionosphere's density is both highly variable and can extend to significant altitudes, exceeding $10^3$ cm$^{-3}$ well above 10 km. However, the interpretation of the Luna data is model dependent, as Bauer (1996) concluded that the Luna data were consistent with no significant lunar ionosphere.

In addition to an ion or molecular component to the plasma layer above the Moon's surface, there are reports of a "horizon glow" from both crewed and robotic missions (e.g. Rennilson and Crisswell 1974; Zook and McCoy 1991). There is widespread agreement that this "horizon glow" is likely due to electrostatically charged dust that is levitated above the surface. Such a component would also contribute free electrons to the atmosphere.

More recently, there have been a series of spacecraft-based remote sensing efforts to measure the lunar ionosphere. Pluchino et al. (2008) performed dual-frequency (2200 and 8400 MHz) lunar occultation observations of the SMART-1, *Cassini*, and Venus Express spacecraft. One of the experiments on the Japanese SELENE (KAGUYA) mission used a series of dual-frequency measurements (at 2200 and 8500 MHz) in an effort to detect the lunar ionosphere (Imamura et al. 2008). In general, an increase in the electron density on the solar illuminated side of the Moon has been observed, consistent with the expectations for the presence of a lunar ionosphere.

The principle underlying relative ionospheric opacity measurements or riometry is that the refractive index of a fully or partially ionized medium (plasma) is a function of frequency and becomes negative below a characteristic frequency known as the *plasma frequency*. At frequencies below the plasma frequency, an electromagnetic wave cannot propagate through the medium and is reflected upon incidence. The plasma frequency is given by $\omega_p = (4\pi n_e e^2/m_e)^{1/2}$ where $n_e$ is the electron density, e is the charge on the electron, and $m_e$ is the mass of the electron. With $\omega_p = 2\pi \nu_p$, and substituting for physical constants, $\nu_p = 9$ kHz $(n_e/1 \text{ cm}^{-3})^{1/2}$.

A riometer exploits this characteristic of a plasma to measure the ionospheric density. If a broadband reference emitter with a known spectrum is observed through the ionosphere, the ionosphere's plasma frequency, and in turn the (peak) ionospheric density, can be determined from the frequency at which absorption occurs. By monitoring the plasma frequency, a riometer can track changes in the ionospheric density over time. In practice, the most commonly used reference emitter is the non-thermal radio emission from the Milky Way Galaxy. This Galactic emission has the favorable properties of being both extremely well characterized and constant in time.

Riometers have been used for decades, in remote and hostile environments, for tracking the properties of the Earth's ionosphere. Only a few antennas are needed for a riometer, and deploying such would serve as an initial demonstration of the technologies en route to the deployment of a larger array. In addition to basic lunar science, riometer measurements would support lunar exploration by tracking the modification of the lunar atmosphere by exhaust from the sample return rockets and future landers.

### *4.2.3 Measurement of Interplanetary Nanodust*

Interplanetary space is pervaded by dust with sizes ranging from nanometers to tens of microns and larger. Recent work on interplanetary dust has revealed a substantial population of nanometer-size dust, or nanodust, with fluxes hundreds of thousands of times higher than better understood micron-sized dust grains. This nanodust tends to move with the speed of the solar wind, or at hundreds of km/sec, as opposed to more typical Keplerian speeds of tens of km/sec. Since impact damage grows faster than the square of the impact speed for high speed dust, this nanodust can generate significant damage when it impacts an object such as a spacecraft, or a planet, or the Moon, or an asteroid. In this section, we describe how a low frequency radio array is ideal for measuring the distribution of dust particles as a function of size in interplanetary space, and ultimately for understanding how dust modifies the surfaces of planets and other objects in the solar system.

Dust has many sources, including collisions between asteroids, escaping gas from comets, and condensation within the solar atmosphere. Additional dust streams into the solar system from interstellar space. The size, speed, and mass distribution of dust in interplanetary space and its variation with time tell us about the history of these sources. Measurements of dust properties have been performed with dedicated dust instruments specifically designed to characterize dust particles (Grun 1993). More recently, it has been shown that space-based radio receivers can also be used to measure dust. These radio instruments function by measuring the electrical signals produced when dust grains impact objects at high speed and create expanding clouds of plasma (e.g., Meyer-Vemet et al. 2009; Zaslavsky et al. 2012; and references therein). Work on the use of radio receivers for studying dust have shown that radio observations have two particular strengths when it comes to conducting a survey of the interplanetary dust population. First, radio arrays are very sensitive to nanodust, a major fraction of the dust population in the solar system that produces weak signals in standard instruments. Second, a radio array is also ideal for searching for the highest mass, but rarest dust particles. This is because the entire surface area of the array, which for lunar concepts would exceed thousands of square meters, becomes a single sensitive dust detector.

Recently, Meyer-Vernet et al. (2009) proposed that small signals seen by the electric field antennas on one of the STEREO spacecraft may be due to nanometer scale dust particles. In order to determine the range of masses, and the uniqueness of radio measurements of dust properties, Zaslavsky et al. (2012) analyzed dust impacts recorded by the STEREO/WAVES radio instrument onboard the two STEREO spacecraft near 1 astronomical unit (AU) during the period 2007-2010.

The impact of a dust particle on a spacecraft produces a plasma cloud whose associated electric field can be detected by on-board electric antennas (Fig. 12). When a grain impacts an object at extremely high speeds (a hyperkinetic impact), it generates a cloud of high temperature plasma of total charge approximately proportional to $mv^{3.5}$, where m is the mass of the grain and v is the speed. Since ions and electrons in the cloud have the same thermal energy, the electrons expand much more quickly, and create a large potential drop along the antenna. Analysis suggests that this technique works very well for measurements that cover the mass intervals $10^{-22} - 10^{-20}$ kg and $10^{-17} - 5\times10^{-16}$ kg (Zaslavsky et al. 2012). The flux of the larger dust agrees with measurements of other instruments on different spacecraft, and the flux of the smaller dust grains agrees with theoretical predictions.

For a lunar radio array with 3 arms of 500 m length each and average antenna width on the arms of 1 m, the surface area would be 1500 $m^2$. Given the expected flux distribution, this would correspond to approximately 1000 impacts/sec for nanodust, and detections of the heavy 10 micron dust several times a minute.

## 5. Summary and Conclusions

The lunar farside is a huge, unexplored "new world" in Earth's backyard. It is dramatically different from regions of the Moon investigated by Apollo containing, for example, only ~1% maria versus 31% for the nearside. The farside includes the South Pole-Aitken basin (SPA), the largest and deepest basin on the Moon, and possibly the oldest impact site in the inner solar system. Because of the Moon's tidal locking with respect to the Earth, the farside may be the only known radio-quiet site that could be used to probe the faint, low radio frequency signals coming from the Dark Ages and Cosmic Dawn of the early Universe.

We have proposed a novel human/robotic mission concept that would be the first to explore the farside and perform high priority science from its surface. The Orion MPCV, launched by NASA's SLS, would be placed into a halo orbit about the Earth-Moon Lagrange point L2. Such an orbit would place Orion astronauts in a unique position to see both the lunar farside and to have direct communications with Earth.

A separate unmanned robotic craft is proposed to land inside the smaller Schrödinger basin, which itself is within the SPA. Schrödinger has the unique attribute of being one of the youngest impact basins on the Moon but within the oldest basin. We propose the first telerobotic exploration of this basin, controlled by astronauts aboard Orion. A sample return from the SPA was identified as a high priority for the *Vision and Voyages for Planetary Science in the Decade 2013-2022* report (NRC, 2011). It would be a powerful test of the lunar cataclysm and lunar magma ocean hypotheses, two of the most fundamental ideas that emerged from the Apollo program.

We also propose to take advantage of the unique radio-quiet zone of the farside to deploy a polyimide film low radio frequency array using telepresence. This array is designed to detect highly redshifted 21-cm signals from the early Universe's Cosmic Dawn which would be the first measure of how and when did the first objects (stars and quasars) "light up" the Universe – a top science objective of the *New Worlds, New Horizons in Astronomy and Astrophysics Decadal Survey* (NRC, 2010). This array would also be the first to definitively measure the density, and its diurnal variations, of the ionized lunar atmosphere. Finally, it could also serve as a powerful detector of high velocity nanodust which may be the major force in weathering airless bodies such as the Moon and asteroids.

This proposed mission is also an opportunity to validate the use of sensors on a deep space mission for the real-time monitoring of ionizing radiation hazards from galactic cosmic rays and solar energetic particles. A range of instruments are possible, from sophisticated particle spectrometers that measure the distribution of linear energy transfer (LET) through simulated human tissue (Spence et al. 2010) to compact micro-dosimeters (Mazur et al. 2011) that track total dose in an extremely small resource volume. The examples described here are currently in orbit on the NASA Lunar Reconnaissance Orbiter spacecraft. Flying the spares of these instruments on the L2-Farside mission would permit comparison of simultaneous measurements of radiation dose near the surface of the Moon and in deep space.

We emphasize that an important driver for this mission will be to demonstrate realistic proof-of-concept exploration strategies that will be needed for early missions to asteroids and to the Mars system (including Deimos and Phobos). Unlike Apollo-style sortie missions, the L2-Farside human/robot concept does not require a human-rated lander and is thus affordable within NASA's current notional budgets for the remainder of this decade.

In summary, the proposed L2-Farside mission would offer a number of science and exploration "firsts." It would be the first mission to the surface of the Moon's farside. It would be the first to investigate and potentially return samples from the oldest impact basin in the inner solar system, possibly holding a key to understanding the formation and evolution of the Earth-Moon system. The L2-Farside mission would deploy a unique polyimide film low radio frequency array in a proven radio-quiet zone which could be the first to observe the currently unexplored epoch of the Universe when the first stars and galaxies formed. Finally, this mission would be the first to demonstrate teleoperation of rovers by astronauts in orbit to undertake geological explorations and to collect samples as well as to deploy sophisticated scientific instrumentation on an extraterrestrial body.

**Acknowledgements**

The development of this mission concept was supported by the Lunar University Network for Astrophysics Research (http://lunar.colorado.edu), headquartered at the University of Colorado Boulder, and the LPI-JSC Center for Lunar Science and Exploration in Houston (http://www.lpi.usra.edu/nlsi/), both funded by the NASA Lunar Science Institute (NASA Cooperative Agreements NNA09DB30A and NNA09DB33A, respectively). Part of this research was conducted at the Jet Propulsion Laboratory, California Institute of Technology, under contract with NASA. We thank Y. Takahashi for the use of his electromagnetic propagation code to study diffraction effects at locations on the lunar farside and A. Datta for porting/running the code at U. Colorado. We also thank Mattie Toll for her editorial assistance and Terry Fong for insightful discussions on telerobotics In addition, we are grateful to Dan Lester for helpful comments on an earlier draft of this paper.

**References**

Alexander, J. K., Kaiser, M. L., Novaco, J. C., Grena, F. R., & Weber, R. R. Scientific instrumentation of the Radio-Astronomy-Explorer-2 satellite. A&A, 40, 365-371, 1975.

Alkalai, L., Solish, B., Elliott, J. Orion/MoonRise: A Human & Robotic Sample Return Mission from the Lunar South Pole-Aitken Basin, Proceedings IEEE Aerospace Conference, Big Sky Montana, March 2013, in press, 2013.

Bauer, S.J. Limits to a Lunar Ionosphere. Sitzungsberichte und Anzeiger, Abt. 2, 133, 17, 1996.

Bobskill, M., and Lupisella, M. The Role of Cis-Lunar Space in Future Global Exploration. GLEX-2012.05.5.4x12270. In Proceedings of IAF/AIAA Global Space Exploration Conference, Washington, 2012.

Bualat, M. Deans, M., Fong, T., Provencher, C., Schreckenghost, D., and Smith, E. ISS crew control of surface telerobots. GLEX-2012.01.2.7x12188. In Proceedings of IAF/AIAA Global Space Exploration Conference, Washington, 2012.

Burns, J.O., Lazio, J., Bale, S. et al. Probing the first stars and black holes in the early Universe with the Dark Ages Radio Explorer (DARE). Advances in Space Research, 49, 433-450, 2012.

de Oliveira-Costa, A., Tegmark, M., Gaensler, B.M., Jonas, J., Landecker, T.L., Reich, P. A model of diffuse Galactic radio emission from 10 MHz to 100 GHz. MNRAS, 388, 247-260, 2008.

Duke, M.B. Sample Return from the Lunar South Pole-Aitken Basin. Adv. Space Res., 31, 2347-2352, 2003.

Farquhar, R, The Utilization of Halo Orbits in Advanced Lunar Operations, NASA TN D-

6365, July 1971

Fong, T., Berka, R., Bualat, M., Diftler, M., Micire, M., Mittman, D., Sunspiral, V., and Provencher, C. The Human Exploration Telerobotics Project. GLEX-2012.01.2.4x12180. In Proceedings of IAF/AIAA Global Space Exploration Conference, Washington, 2012.

Goddard Space Flight Center (GSFC). Telerobotics Symposium, Greenbelt, MD, http://telerobotics.gsfc.nasa.gov/ , 2012.

Grun, E., Zook, H.A., Baguhl, M. et al. Discovery of Jovian dust streams and interstellar grains by the Ulysses spacecraft. Nature 362, 428-430, 1993.

Hodges, R.R., Hoffman, J.H., & Johnson, F.S. The Lunar Atmosphere. Icarus, 21, 415-426, 1974.

Hopkins, J. Early Telerobotic Exploration of the Lunar Farside Using Orion Spacecraft at Earth-Moon L2. GLEX-2012.02.3.2x12595. In Proceedings of IAF/AIAA Global Space Exploration Conference, Washington, 2012.

Hopkins, J., Bryukhanov, N., Murashko, A., Walther, S., Nakanishi, H., International Cooperation Mission (ICM) Towards Future Space Exploration, GLEX-2012.05.4.5x12592. In Proceedings of IAF/AIAA Global Space Exploration Conference, Washington, 2012.

Imamura, T., Iwata, T., Yamamoto Z., et al. Studying the Lunar Ionosphere with SELENE Radio Science Experiment. American Geophysical Union, #P51D-04, 2008.

Johnson, F. S. Lunar Atmosphere. Rev. Geophys. Space Phys., 9, 813-823, 1971.

Jolliff, B.L., Shearer, C.K., Papanastassiou, D.A., Alkalai, L., and the Moonrise Team. MoonRise: South Pole-Aitken Basin Sample Return Mission for Solar System Science, in Annual Meeting of the Lunar Exploration Analysis Group, Abstract #3072, p. 31. LPI Contribution No. 1595, Lunar and Planetary Institute, Houston, TX, 2010

Kramer, G. Y., Kring, D. A., Nahm, A. L., Pieters, C. M. Spectral and photogeologic mapping of Schrödiniger basin and implications for Post-South Pole-Aitken impact deep subsurface stratigraphy. Icarus, submitted, 2012.

Kring, D. A. Environmental consequences of impact cratering events as a function of ambient conditions on Earth. Astrobiology 3, 133-152, 2003.

Kring, D. A. What can astronauts learn from terrestrial impact craters for operations on the Moon and Mars? Nördlingen 2010: The Ries Crater, the Moon, and the Future of Human Space Exploration, Abstract #7036, http://www.lpi.usra.edu/meetings/nordlingen2010/pdf/7036.pdf, 2010.

Kwong, J., Norris, S. D., Hopkins, J. B., Buxton, C. J., Pratt, W. D., Jones, M. R. Stepping Stones: Exploring a Series of Increasingly Challenging Destinations on the Way to Mars. AIAA Space 2011 Conference. Long Beach, CA. 27 September 2011 to 29 September 2011.

Landis, Geoffrey A., Footsteps to Mars: An Incremental Approach to Mars Exploration Journal of the British Interplanetary Society, Vol 48, pp 367-342, 1995.

Lazio, T.J.W., MacDowall, R.J., Burns, J. O., Jones, D. L., Weiler, K. W., Demaio, L., Cohen, A., Paravastu, N., Dalal, Polisensky, E., Stewart, K., Bale, S., Gopalswamy, N. Kaiser, M., Kasper, J. The Radio Observatory on the Lunar Surface for Solar studies. Advances in Space Research, 48, 1942-1957, 2011.

Lester, D.F. and Thronson, H.A. Human space exploration and human spaceflight: Latency and the cognitive scale of the universe. Space Policy, 27, 89-93, 2011a.

Lester, D. F. and Thronson, H. A. Low-Latency Lunar Surface Telerobotics from Earth-Moon

  Libration Points. American Institute of Aeronautics and Astronautics Space 2011, AIAA-2011-7341, 2011b.

Lunar Exploration Analysis Group (LEAG). Lunar Exploration Roadmap v.1-1. http://www.lpi.usra.edu/leag/ler_draft.shtml, 2011.

Mazur, J. E., Crain, W.R., Looper, M.D., Mabry, D. J., Blake, B., Case, A. W., Golightly, M. J., Kasper, J. C., and Spence, H. E. New measurements of total ionizing dose in the lunar environment, Space Weather, 9, S07002, doi:10.1029/2010SW000641, 2011.

Mest, S. C. The geology of Schrödinger basin: Insights from post-Lunar Orbiter data. Recent Advances in Lunar Stratigraphy. Williams, D.A., and Ambrose, W. (eds.), Geological Society of America Special Paper 477, pp. 95–115, Boulder, CO, 2011.

Meyer-Vernet, N., Maksimovic, M., Czechowski, A. et al. Dust detection by the wave instrument on stereo: nanoparticles picked up by the solar wind? Solar Physics, 256, 463-474, 2009.

National Research Council. The scientific context for exploration of the Moon. National Academy Press, Washington, 107pp, 2007.

National Research Council, Committee on the Evaluation of Radiation Shielding for Space Exploration, National Research Council, Managing Space Radiation Risk in the New Era of Space Exploration. National Academy Press, Washington, 132pp, 2008.

National Research Council, Committee for a Decadal Survey of Astronomy and Astrophysics, New Worlds, New Horizons in Astronomy and Astrophysics, National Academies Press, Washington, 324pp, 2010.

National Research Council, Committee on the Planetary Science Decadal Survey, Vision and Voyages for Planetary Science in the Decade 2013-2022, National Academies Press, Washington, 382pp, 2011.

O'Sullivan, K.M., Kohout, T., Thaisen, K. G., and Kring, D. A. Calibrating several key lunar stratigraphic units representing 4 billion years of lunar history within Schrödinger basin. Recent Advances in Lunar Stratigraphy, Williams, D. A. and Ambrose, W. (eds.), pp. 117–128, Geological Society of America Special Paper 477, Boulder, CO, 2011.

Pluchino, S., Schillirò, F., Salerno, E., Pupillo, G. Maccaferri, G. & Cassaro, P. Radio Occultation Measurements of the Lunar Ionosphere. Memorie Soc. Astron. Italiana Suppl., 12, 53-59, 2008.

Pritchard, J. & A. Loeb, A. 21-cm Cosmology in the 21$^{st}$ Century. Reports on Progress in Physics, 75, 086901, 35pp, 2012.

Reasoner, D.L., & O'Brien, B. J. Measurement on the lunar surface of impact-produced plasma clouds. J. Geophys. Res., 77, 1292-1299, 1972.

Rennilson, J. J. & Criswell, D. R. Surveyor Observations of Lunar Horizon-Glow. Moon, 10, 121-142, 1974.

Schmidt, G. R., Landis, G. A., and Oleson, S. R. HERRO Missions to Mars and Venus Using Telerobotic Surface Exploration From Orbit. International Astronautical Congress, Cape Town, South Africa, 2011.

Shoemaker, E. M., Robinson, M. S., and Eliason, E. M. The South Pole Region of the Moon as Seen by Clementine. Science, 266, 1851-1854, 1994.

Singer, S. Fred, The Ph-D Proposal: A Manned Mission to Phobos and Deimos, Procceedings of The Case for Mars Conference, Boulder, CO, April 29-May 2, 1981, Univelt Inc, San Diego, CA, p 39-65, 1984.

Spence, H.E., Case, A.W., Golightly, M.J. et al. CRaTER: The Cosmic Ray Telescope for the Effects of Radiation Experiment on the Lunar Reconnaissance Orbiter Mission, SPACE SCIENCE REVIEWS, Volume 150, Numbers 1-4, 243-284, DOI: 10.1007/s11214-009-9584-8, 2010.

Takahashi, Y. A Lunar Far Side Radio Array As The First Astronomical Observatory On The Moon: Precursor Studies. EGS XXVII General Assembly, Nice, 21-26 April 2002, abstract #5174, 2002.

Vyshlov, A. S., & Savich, N. A. Observations of radio source occultations by the Moon and the nature of the plasma near the Moon. Cosmic Res., 16 (transl. Kosmicheskie Issledovaniya, 16), 551-556, 1978.

Wilhelms, D. E., McCauley, J. F., Trask N. J. The geologic history of the Moon. U.S. Geological Survey, Washington : U.S. G.P.O. ; Denver, CO (Federal Center, Box 25425, Denver 80225), 302pp, 1987.

Yamamoto, S., Nakamura, R., Matsunaga, T., Ogawa, Y., Ishihara, Y., Morota, T., Hirata, N., Ohtake, M., Hiroi, T., Yokota, Y., and Haruyama, J. Olivine-rich exposures in the South Pole-Aitken basin. Icarus, 218, 331-344, 2012.

Zaslavsky, A., Meyer-Vernet, N., Mann, I. et al. Interplanetary dust detection by radio antennas: Mass calibration and fluxes measured by STEREO/WAVES. Journal of Geophysical Research, 117, A05102, 13pp, 2012.

Zook, H.A. & McCoy, J.E. Large scale lunar horizon glow and a high altitude lunar dust exosphere. Geophys. Res. Lett., 18, 2117-2120, 1991.

**Figures**

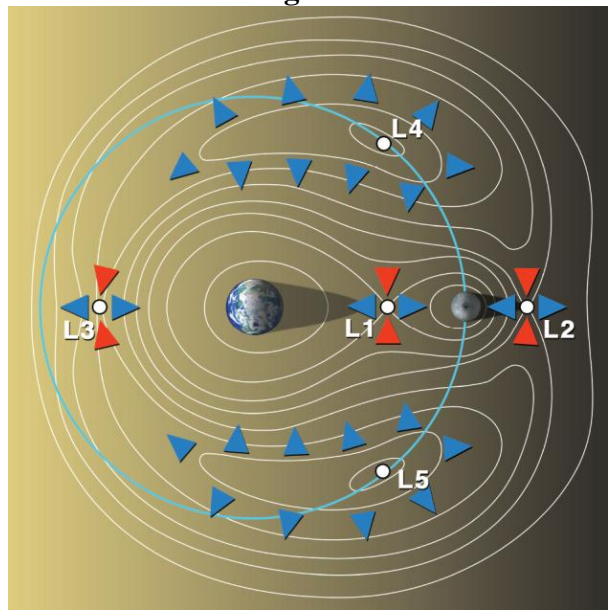

Fig. 1: Earth-Moon Lagrange Points, all of which are gravitational equilibrium points in the Sun-Earth-Moon system. The contours illustrate gravitational equipotential surfaces. Like elevation in a topographic map, the potential in this illustration is the same along any contour and gradients in the effective potential are greatest when the contours are close together. Red and blue arrows indicate the directions an object will tend to wander in the vicinity of the Lagrange points. We propose an early mission with Orion placed in a halo orbit at Earth-Moon L2. Astronauts would teleoperate robots on the lunar farside to gather samples from ancient impact craters and deploy a low frequency array of radio antennas to track the effects of the first stars and galaxies in the early Universe.

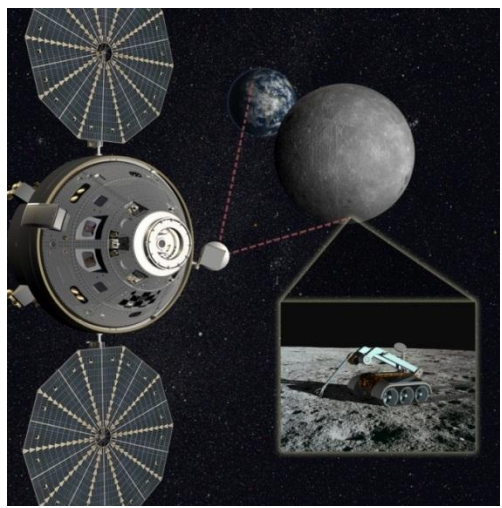

Fig. 2: Schematic picture of the Orion MPCV L2 mission in which astronauts teleoperate a rover on the lunar farside.

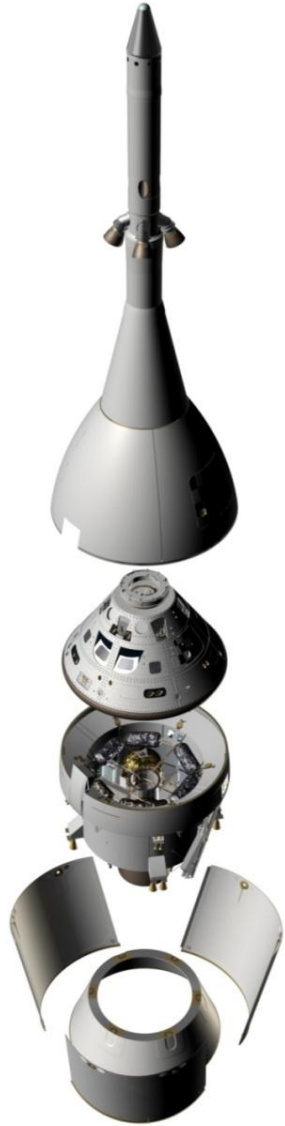

Fig. 3: Orion MPCV spacecraft. From top to bottom, the Launch Abort System (LAS), Command Module (CM), Service Module (SM), and Spacecraft Adapter. See text in Section 2.1 for further details.

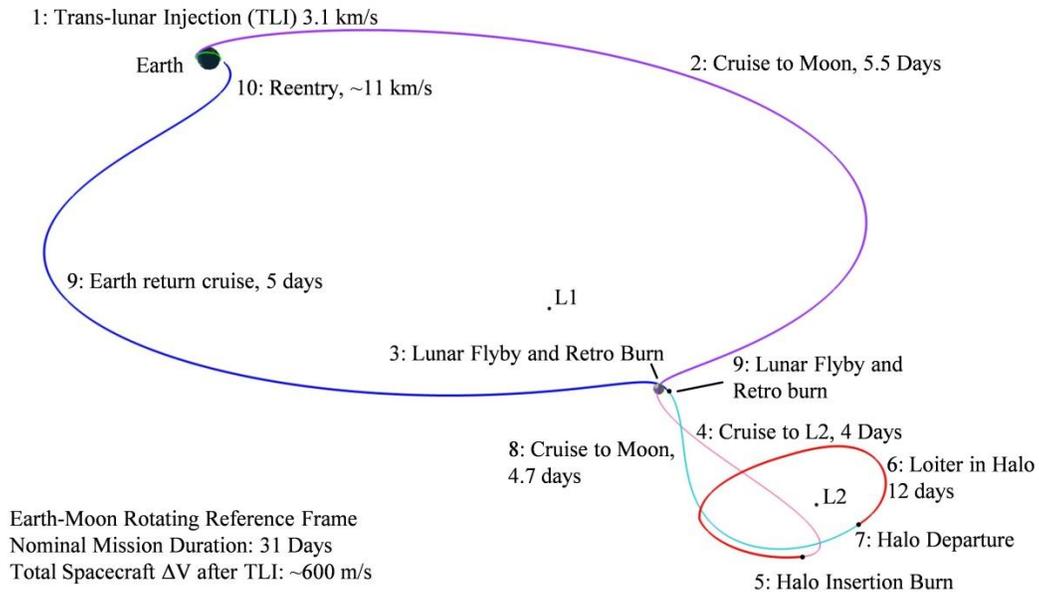

Fig. 4: The crew segment of the L2-Farside mission can be performed using a single SLS, along with the Orion MPCV in about 30 days.

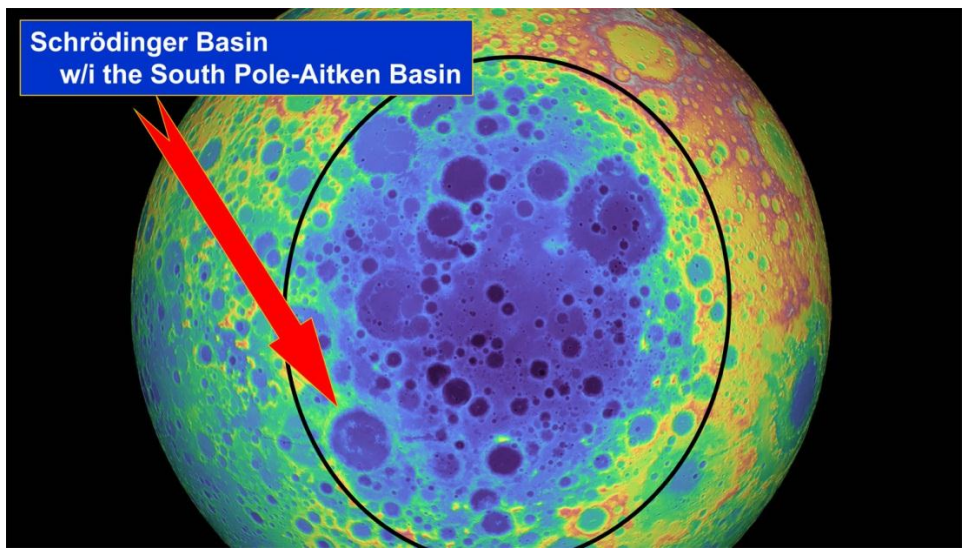

Fig. 5: False-colored topography of the southern lunar farside highlighting the 13 km deep South Pole-Aitken basin (blue colors) and the surrounding highlands (yellows and reds). Schrödinger basin (also blue) is marked on this image and lies within 500 km of Shackleton crater at the lunar South Pole. Background topography courtesy of LRO-LOLA/NASA GSFC SVS.

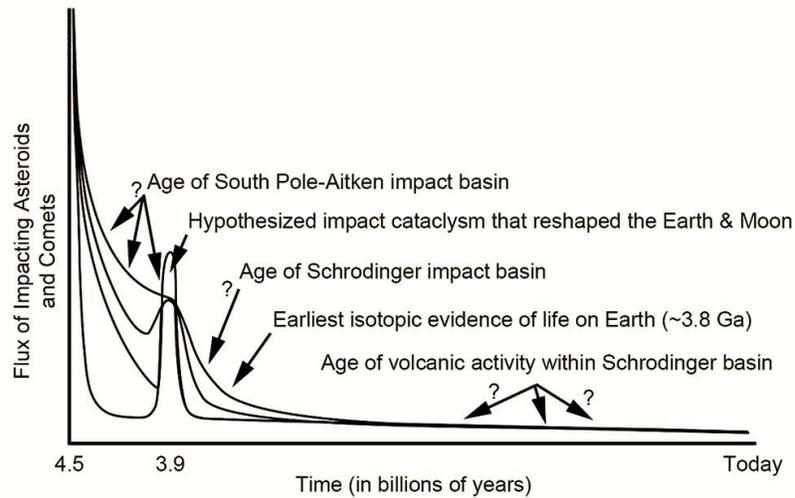

Fig. 6 : Schematic diagram illustrating the evolution of the impact flux to the Earth and Moon as a function of time from 4.5 billion years ago to the present day. Following the accretion of the Earth and Moon, the impact flux declined at a still unknown rate (illustrated with multiple curves). Apollo samples suggest there was a surge in the impact rate circa 4 billion years ago. The South Pole-Aitken basin represents the oldest basin-forming event, and the Schrödinger basin represents the second youngest basin-forming event. Soon after that period of bombardment, the earliest isotopic evidence of life on Earth appears. Diagram modified from Kring (2003).

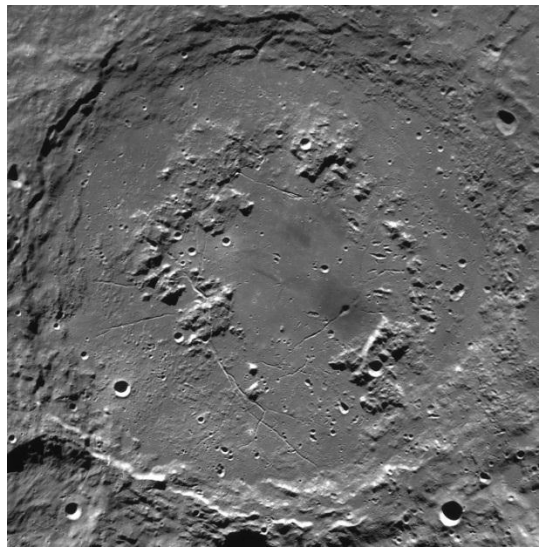

Fig. 7. An image of the ~320 km diameter Schrödinger basin. The relatively flat floor is covered with solidified impact melt that could be sampled to determine the age of the basin. An uplifted peak provides outcrops of rock from the lunar magma ocean. Two dark-colored regions within the crater are volcanic deposits that erupted more recently and provide evidence of magmatic processes deep within the lunar interior.

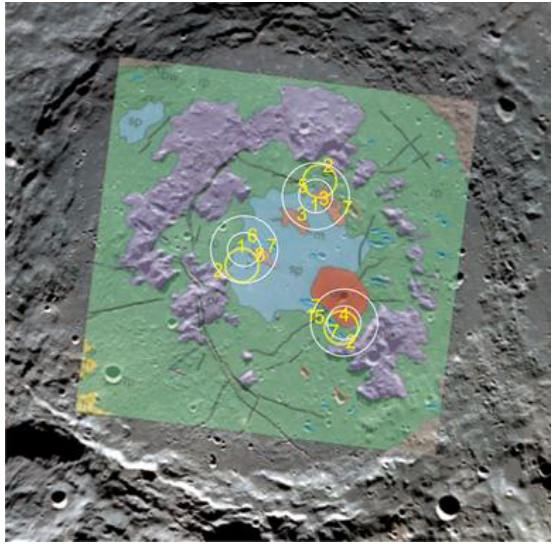

Fig. 8 : Three regions provided multiple potential landing sites and stations, as developed by O'Sullivan et al. (2011). The landing sites would provide access to impact melt lithologies (in all cases), uplifted samples of the lunar magma ocean (in all cases), mare volcanism (in one and possibly two cases), and pyroclastic volcanism (in one case). Yellow circles represent 10 km diameter region in which all lithologies exist. White circles, centered around prefered landing sites, represent 10 and 20 km radial distances. Geologic units were mapped by Shoemaker et al. (1994).

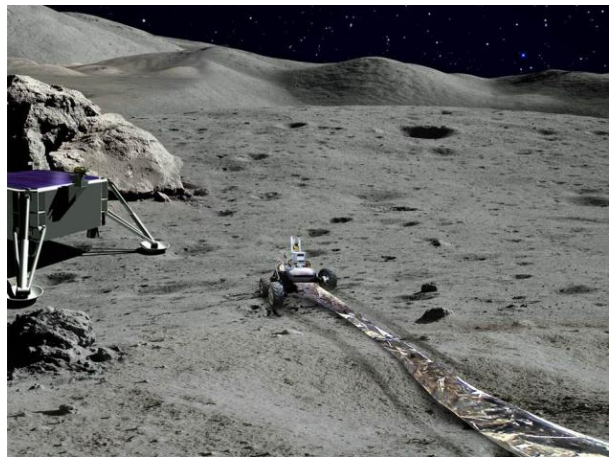

Fig. 9: Artist's impression of a telerobotically-operated rover deploying a polyimide film-based antenna. The lander that would carry the rover and antenna array to the surface is visible to the left.

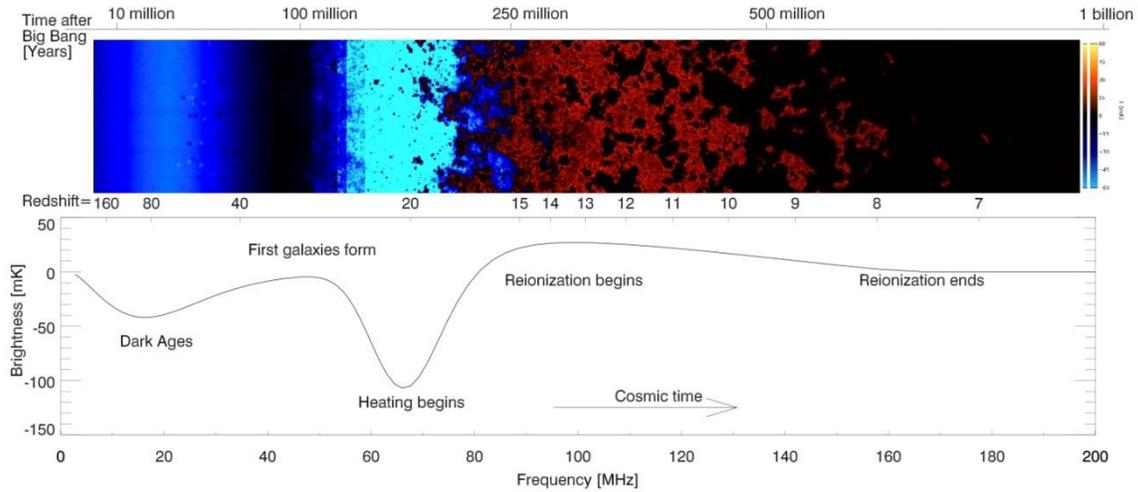

Fig. 10: (*Top*) Time evolution of the 21-cm brightness from just before the first stars form (Dark Ages) through to the end of the EoR. Color indicates the strength of the 21-cm brightness as it transitions from absorption (blue) to emission (red) and finally disappears (black) due to ionization. (*Bottom*) Expected evolution of the sky-averaged 21 cm brightness from the "Dark Ages" at $z = 150$ to the end of the EoR sometime before $z = 6$. The frequency structure is driven by the interplay of gas heating, the coupling of gas and 21 cm temperatures, and the ionization of the gas. Figure is courtesy of Jonathan Pritchard (see Pritchard & Loeb, 2012).

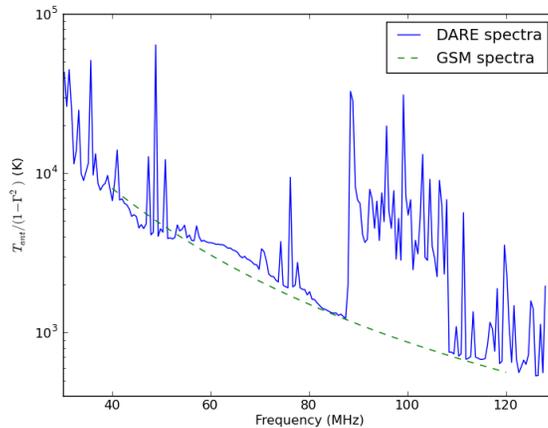

Fig. 11: Recently measured radio spectrum over most of the low frequency range relevant for Cosmic Dawn observations. The ordinate axis is antenna temperature (defined via $P=kT_{ant}\Delta\nu$, where P is the power measured by the radio receiver, k is Boltzmann's constant, and $\Delta\nu$ is the frequency bandwidth), corrected for the instrumental response ($\Gamma$ is the antenna reflection coefficient). This spectrum was acquired in the National Radio Quiet Zone surrounding the Green Bank site of the National Radio Astronomy Observatory using the Dark Ages Radio Explorer (DARE) engineering prototype (Burns et al. 2012). Note the saturation by RFI in the FM radio band (88–108 MHz). The undulating baseline at frequencies below about 80 MHz is due partially to the effects of the Earth's ionosphere. The green dashed line (GSM = Global Sky Model derived from de Oliveira-Costa et al. 2008) is the expected radio emission from the Milky Way.

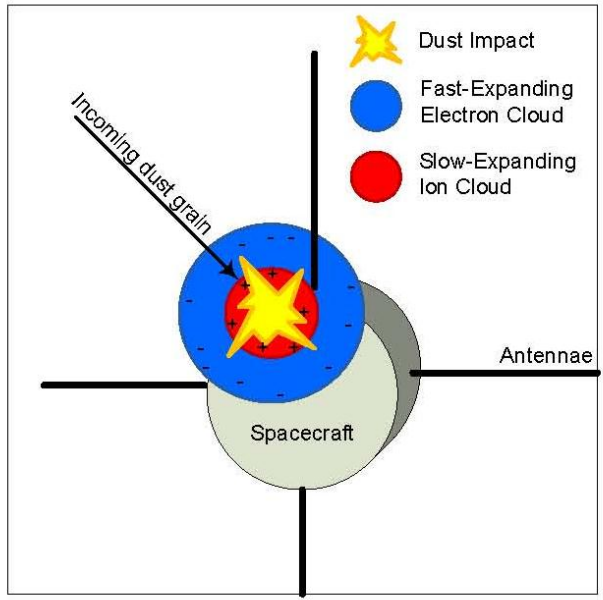

Fig. 12: Creation of an expanding cloud of plasma surrounding a spacecraft after the impact of a dust grain. Since ions and electrons in the cloud have the same thermal energy, the electrons expand much more quickly.

Table 1
Orion ECLSS architecture.

| Function | Technology |
|---|---|
| $CO_2$ and $H_2O$ control | Solid amine bed |
| O2 provision & pressure control | O2 and N2 tanks with total pressure and oxygen sensors |
| Fire detection & suppression | Smoke detectors and nitrogen suppression |
| Food provision | Prepackaged food |
| Waste management | Toilet with fecal collection containers and urine venting |
| Water management | Water tanks |